\documentstyle[prl,aps,twocolumn]{revtex}
\begin{document}
\draft
\title{
Magic Numbers and Optical Absorption Spectrum in Vertically Coupled 
Quantum Dots in the Fractional Quantum Hall Regime
}
\author{Hiroshi Imamura, Peter A. Maksym$^{(a)}$ and Hideo Aoki}
\address{Department of Physics, University of Tokyo, 
Hongo, Tokyo 113, Japan \\ 
$(a)$ Department of Physics and Astronomy,
University of Leicester, Leicester LE1 7RH, United Kingdom}
\date{\today}

\maketitle
\begin{abstract}
Exact diagonalization is used to study the quantum 
states of vertically coupled quantum dots in strong magnetic
fields. We find a new sequence of angular momentum magic numbers
which are a consequence of the electron correlation in the double
dot. The new sequence occurs at low angular momenta and changes into
the single dot sequence at a critical angular momentum determined by
the strength of the inter-dot electron tunneling. 
We also propose that the magic numbers can be investigated 
experimentally in vertically coupled dots. Because of the
generalized Kohn theorem, the far-infrared optical absorption
spectrum of a single dot is unaffected by correlation but the theorem
does not hold for two vertically coupled dots which have different
confining potentials.
We show that the absorption energy of the double dot should exhibit 
discontinuities at the magnetic fields where the total angular
momentum changes from one magic number to another.
\end{abstract} 
\narrowtext
Low-dimensional confined electronic systems, such as quantum wells,
wires and dots, have recently attracted much interest 
because they exhibit dramatic quantum effects when they are placed in
a strong magnetic field. The fractional quantum Hall (FQH) effect 
\cite{fqhe}, which occurs in a two dimensional sheet of electrons, is
probably the most spectacular example although there are many others.
Recently, attention has focused on double-layer FQH 
systems\cite{moon,nakajima,eisenstein,suen,yoshioka2}, where 
the additional degree of freedom (a pseudo-spin that labels the
layers) 
enriches the physics. A central issue in these systems is the
interplay 
of electron correlation and inter-layer electron tunneling --- 
because of the competition between these effects the quantum Hall
state evolves 
continuously from a correlation-dominated (two-component) state
down to a tunneling-dominated (single-component) state within the
quantum Hall regime.

Another recent development is the study of laterally confined
systems, and quantum dots in particular. A dot in the FQH regime
contains a few electrons confined on a length scale of the order of
the magnetic length.
As a function of the total angular momentum, $L$, the ground-state
energy 
of this system exhibits downward 
cusps at specific $L$ values which are known as the 
"magic numbers" \cite{girvinjack,maksymprl}. 
The magic numbers are caused by electron correlation, and there are 
general arguments that relate the magic numbers for given numbers of 
electrons, $N_e$, to the symmetry of the wavefunction and the 
requirement that it satisfies Pauli's principle 
\cite{maksymphysica,maksymprb95,ruan}.
For instance, the magic numbers, $L=3, 6, 9, \cdots$, 
for three spin-polarized electrons 
correspond to triangular spatial correlation that 
minimizes the Coulomb repulsion.

It is then intriguing to ask what happens if 
we laterally confine a double-layer FQH system to form 
vertically-coupled quantum dots.  
Because fascinating correlation effects are known to occur in double
2D 
systems (bilayers) and double 1D systems (double quantum wires 
\cite{takashi}), we can expect to find interesting phenomena in 
double 0D systems (double dots) which are the subject of the present 
work. Specific questions we address are, 
firstly, what will happen to the magic numbers as we vary the
strength of 
the tunneling, and secondly, whether the magic numbers may become 
observable in double dots.  
Technically, we believe that the structures 
considered here are within the scope of current fabrication
technology \cite{Heitmann}.

The system we study is a double dot containing a total of three 
spin-polarized 
electrons. In both dots the electron motion is perfectly two
dimensional 
and the lateral confining potential within each layer 
is assumed to be parabolic. The dots are separated in the vertical 
direction with their centers aligned on a common axis. 
The electrons experience both intra- and inter-layer 
Coulomb repulsions in the presence of the inter-layer 
tunneling. We find a new series of magic numbers, 
which correspond to ground states dominated by the inter-layer 
electron correlation. Very recently, similar effects have been found 
in $1/r^2$ interaction coupled multiple dots in the absence of
tunneling \cite{Benjamin95}.
In addition, we propose that the magic numbers can
manifest themselves in the far-infrared optical absorption
spectrum. In 
a single dot this is impossible because of the generalized Kohn
theorem 
\cite{maksymprl,kohn,johnson89} but the theorem does not hold for a
double 
dot with different confining potentials. Consequently, we find that 
the absorption energy should exhibit discontinuities at the
magnetic fields where the total angular momentum changes from one
magic number to another.


A vertically-coupled double dot is characterized by 
the strength of the parabolic
confinement potential of the upper-(lower-)layer, $\hbar\omega_{+} 
(\hbar\omega_{-})$, 
the layer separation, $d$, and the 
strength of the interlayer tunneling 
(measured by $\Delta_{\rm SAS}$, the energy gap between the symmetric
and antisymmetric states in the non-interacting system).  
The Hamiltonian,
\begin{equation}
  {\cal H} = {\cal H}_{\rm s} + {\cal H}_{\rm t} + {\cal H}_{\rm C},
\end{equation}
comprises the 
single-electron part, ${\cal H}_{\rm s}$, 
the tunneling term, ${\cal H}_{\rm t}$, 
and the Coulomb interaction, ${\cal H}_{\rm C}$. We assume that the 
magnetic field $B$ is so strong that Landau level mixing is
negligible and we write the Hamiltonian in second-quantized form 
with a Fock-Darwin basis\cite{FockDarwin}. This gives

\begin{equation}
  {\cal H}_{\rm s} = \sum_{\ell}\sum_{\alpha}
  \varepsilon_{\ell \alpha}
  c_{\ell \alpha}^{\dag}c_{\ell \alpha},
\end{equation}

\begin{equation}
  {\cal H}_{\rm t} = -\frac{\Delta_{\rm SAS}}{2}\sum_{\ell}
  \left(
    c_{\ell +}^{\dag}c_{\ell -} + c_{\ell -}^{\dag}c_{\ell +}
  \right) ,
\end{equation}

\begin{eqnarray}
  {\cal H}_{\rm c} &=&
  \frac{1}{2}\sum_{\ell_{1}\sim\ell_{4}}\sum_{\alpha_{1}
    \sim\alpha_{4}}
  \langle \ell_{1}\alpha_{1},\ell_{2}\alpha_{2} |
  \frac{e^{2}}
  {\epsilon \sqrt{|{\bf r}_{1}-{\bf r}_{2}|^2 + d^2 }}
  |\ell_{3}\alpha_{3},\ell_{4}\alpha_{4}\rangle\nonumber\\
  &\times&  c_{\ell_{1}\alpha_{1}}^{\dag}
  c_{\ell_{2}\alpha_{2}}^{\dag}
  c_{\ell_{4}\alpha_{4}} c_{\ell_{3}\alpha_{3}}.
\end{eqnarray}
Here, the index $\alpha$ is used to distinguish the two dots, 
$\alpha =+,-$, 
$c_{\ell \alpha}^{\dag} (c_{\ell \alpha})$ are creation 
(annihilation) operators and $\epsilon$ is the dielectric constant
of the host material. The sum over  $\alpha_{1},...,\alpha_{4}$ in 
${\cal H}_{\rm c}$ guarantees that both intra- and inter-layer 
interactions are included.
The energy of the zeroth Landau-level
Fock-Darwin state with angular momentum $\ell$ ($\ge 0$) in the 
$\alpha$\/th dot is $\varepsilon_{\ell \alpha} = (1 + {\ell}) \hbar
(\omega_{\rm c}^{2}/4 + \omega_{\alpha}^{2})^{\frac{1}{2}}
-\ell \hbar\omega_{\rm c}/2$, where $\omega_{\rm c} = eB/m^* c$
is the cyclotron frequency. 

We now estimate typical values of the parameters. 
To obtain approximate values of 
the confinement energy we use a simple electrostatic model in which  
there is a disk of positive charge above and below the two dots,
with the entire structure sandwiched between two metallic disks. This
is meant to mimic the electrostatic confinement scheme in which a
single quantum dot is made by applying a modulated gate electrode to
a modulation doped heterojunction or quantum well \cite{stern}.
For a single dot, we have found that the model is able to reproduce
the confinement energy from an exact solution of the Poisson equation
to about 20\%. To estimate the confinement energy of the double dot
we take typical device dimensions and dopant densities from the work
of Boebinger et al \cite{boeb} and Kumar et al \cite{stern} to find
that $\hbar\omega_\alpha$ is about  2 - 4 meV. 
The asymmetry in $\hbar\omega_\alpha$ depends on the offset of the 
two dots from the symmetric configuration and on the positions of the
disks. It is typically 5-10\% for disk separations of a few hundred 
nanometers and dot offsets of a few tens of nanometers. Larger
asymmetry 
could be achieved by making the structure grossly asymmetric. In our 
calculations we take
$\hbar \omega_+ = 2.0$ meV, $\hbar \omega_- = 2.2$ meV. 
The electrostatic model predicts that the confining potential at 
the center of each dot will be in general different. We assume that
this 
could be compensated by applying a potential to the entire device.  
For the dot separation and
the symmetric-antisymmetric splitting, we take 
typical values from double layer studies of Boebinger et al
\cite{boeb} 
and Eisenstein et al \cite{eisenstein}, leading to $d=20$ nm and 
$\Delta_{\rm SAS}$ in the range 0.2 - 0.5 meV.

The ground-state energy (Fig.~\ref{fig:gene}) is 
calculated as a function of the total angular momentum, $L$, 
by diagonalizing the Hamiltonian 
in a Slater determinant basis 
at $B=10$T for $\Delta_{\rm SAS}=0.2$ meV (solid line) and
$\Delta_{\rm SAS}=0.5$ meV (broken line). 
Qualitatively, the behavior shown in the 
figure is typical for a large range of $B$ values although the value
of 
$L$ at which the minimum energy occurs depends strongly on $B$.
The magic numbers can be 
identified from the positions of downward cusps. 
For $\Delta_{\rm SAS}=0.2$ meV 
we have a new period of two up to $L=3, 5, 7, 9$, 
followed by a period of three, $L=9, 12, \cdots$, 
while for a larger 
$\Delta_{\rm SAS}=0.5$ meV the period is three throughout, 
$L = 3, 6, 9, 12, \cdots$ 
as in the case of a single dot containing three electrons.

To identify the mechanism for the change of period in the magic
number for smaller $\Delta_{\rm SAS}$, 
we show the charge density (the inset to Fig.~\ref{fig:gene}) 
and the pair correlation function (Fig.~\ref{fig:prr0l5}) 
before ($L=5$) and after ($L=12$) the change in the period sets in.  
For $L=5$ (inset (a)) 
the density against the lateral distance from the 
center has a peak
at the center in the lower layer 
while the density is 
double-peaked in the upper layer. 
For $L=12$ (inset (b)) the densities in both layers are 
double-peaked. 
We investigate further by looking at 
the pair correlation function $P({\bf r}, {\bf r}_{0})$ 
(Fig.~\ref{fig:prr0l5}), which is  
defined as the conditional probability of finding an electron at
position ${\bf r}$ given that there is one at position ${\bf r}_0$. 
The fixed electron is at $r_{0} = 16.9$ nm (for $L=5$, a) or 23.5 nm 
(for $L=12$, c) in the upper 
layer where the charge density is a maximum. From the figure 
we can immediately see that the ground state electron configuration 
changes from one dominated by interlayer correlation 
to one dominated by intralayer correlation.  For $L=5$ 
the form of the correlation corresponds to a triangular "electron 
molecule" developing {\it across} the two layers, with  
one electron at the center of the lower layer while 
the other two are in the upper layer. In contrast, the 
triangular form develops within each layer for $L=12$.  
Similar physics should occur when the two dots have the 
same confining potential ($\omega_+ = \omega_-$).

The change in correlation can be understood by considering the
energy. 
As $L$ is decreased the lateral spatial extent of the wave function
 becomes comparable with the vertical separation of the layers.  
When the total angular momentum is small enough, the intralayer
Coulomb
interaction dominates the interlayer Coulomb interaction, 
so electrons try to avoid each other by developing 
an inter-layer correlation.  Although this has to involve mixing of 
states in the two dots and costs an energy 
$\Delta_{\rm SAS}$, the 
electron correlation still dominates as long as 
$\Delta_{\rm SAS}$ is small enough.  
We believe this is why the 
new magic numbers $L=5,7$ appear for smaller $\Delta_{\rm SAS}$.  
The global minimum energy and the angular momentum of the absolute 
ground state depends on the  
magnetic field. By scanning a range of magnetic fields we have found 
that the new magic number states at $L=5$ and $L=7$ become the 
absolute ground state when $B\sim4$T and $B\sim6$T respectively.

A comparison of our results with the phase diagram\cite{qhepd} 
for the bulk double-layer 
FQH system is not straightforward.  
The latter phase diagram 
is drawn against two dimensionless quantities, $d/\ell_B$ and 
$\Delta_{\rm SAS}/(e^2/\epsilon \ell_B)$, 
where $\ell_B=(c\hbar /eB)^{1/2}$ is the 
magnetic length. Because of the confining potential the relevant 
length scale for dots becomes the effective magnetic length
$\lambda$ with $\lambda^2 = \hbar /m^* (\omega_{\rm c}^2 +
4\omega_0^2)^{1/2}$.
With the parameters we have used $\lambda = (0.91\sim 0.97)\ell_B$
for $B=(5\sim 10)$T.  
This yields $e^2/\epsilon \lambda = 14.5$ meV for $B=10$T, so that 
$\Delta_{\rm SAS}/(e^2/\epsilon \lambda) = 0.01 \sim 0.03$ 
for the double dots considered here, 
while the Landau level filling, $\nu$, which is 
usually defined as $\nu = N_e(N_e-1)/2L$ for dots, ranges 
from $\nu=3/5$ for $L=5$ to $\nu=1/4$ for $L=12$. 
It is an interesting problem to see how the intra- to inter-dot
crossover 
in double dots may be related to the one- to 
two-component crossover in the double layers. 

Now we move on to the far-infrared (FIR) optical absorption spectrum.
In a single dot with a parabolic confinement potential, 
the electron-electron interaction does not affect the 
FIR absorption.  This follows from the generalized  
Kohn theorem: 
long-wavelength electromagnetic radiations with electric vector ${\bf
  E}$
couples to the dot via the perturbation Hamiltonian
\begin{equation}
 {\cal H}^{\prime}
  = \sum_{i = 1}^{N} e {\bf E} \cdot {\bf r}_i,
\end{equation}
which depends only on the center-of-mass coordinate. In a single dot 
with parabolic confinement the Hamiltonian separates into 
the center-of-mass and relative (interaction) parts and the latter
is irrelevant to optical transitions. In contrast, the separation
does not occur in vertically coupled dots having different
confinement energies even if both dots have parabolic confinement.

This means the Coulomb interaction should affect 
FIR absorption spectra.  

To quantify the effect we have calculated the 
FIR absorption spectrum of vertically coupled dots 
from the matrix element of the perturbation Hamiltonian,
$\langle {\cal H}^{\prime}\rangle$, between the ground state and all 
the excited states. 
Before discussing the results, we comment on the applicability of
this
approach to 
real systems. One important question is the nature of the electric
field
${\bf E}$.  
Several authors have questioned the relation between the 
applied electric field and the internal electric field in mesoscopic 
systems \cite{Perenboom,Cho,Keller} with the general conclusion that 
depolarization effects are important. Therefore we would have to 
calculate the internal electric field to obtain the absolute value
of the absorption coefficient. In addition, precise calculation of
the absorption spectrum would require us to take account of other
device properties that affect absorption, such as finite thickness of
the individual dots and deviations from a parabolic potential, about
which 
scant information is available. We therefore make the reasonable 
assumption that the internal electric field is uniform and discuss
only 
the absorption energy and the relative intensities of various
transitions. 
This should be sufficient for our purpose of demonstrating 
that the FIR absorption of vertically 
coupled dots is affected by electron correlation.

\begin{figure}
  \caption{FIR absorption spectrum (upper panel) and total angular 
  momentum (lower panel)
    of vertically coupled dots for 
    $\hbar \omega_+ = 2.0$ meV, $\hbar \omega_- = 2.2$ meV, with 
    the layer separation $d = 20$ nm
    and $\Delta_{\rm SAS} = 0.5$ meV. 
    The position of each filled circle gives the energy of the
    transition 
    while the size of the circle represents the absorption 
    intensity. 
    The solid (broken) line corresponds to the 
    single-electron absorption spectrum for
    $\hbar\omega_{\pm}=2.0$ (2.2) meV.  
}
  \label{fig:fir}
\end{figure}
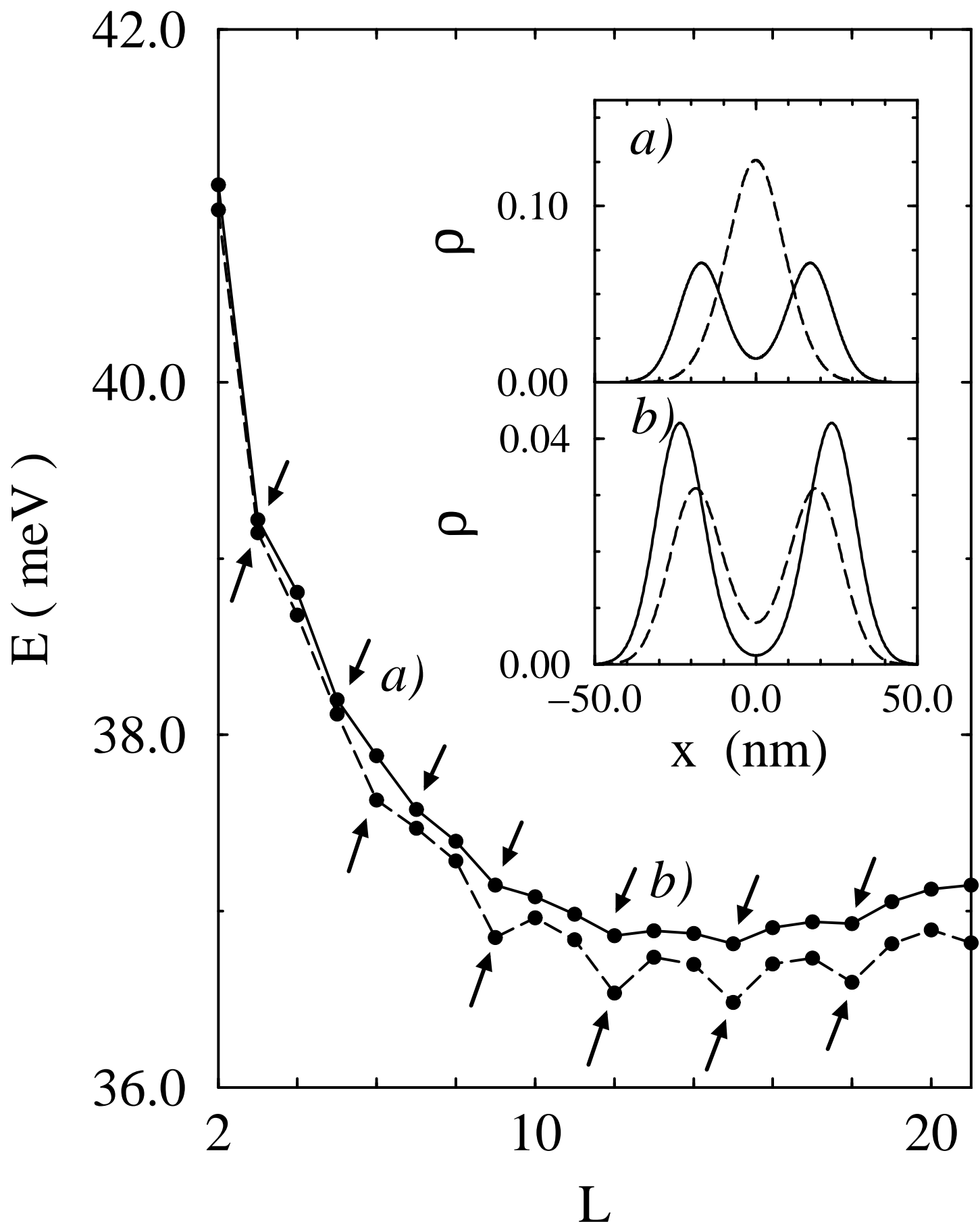
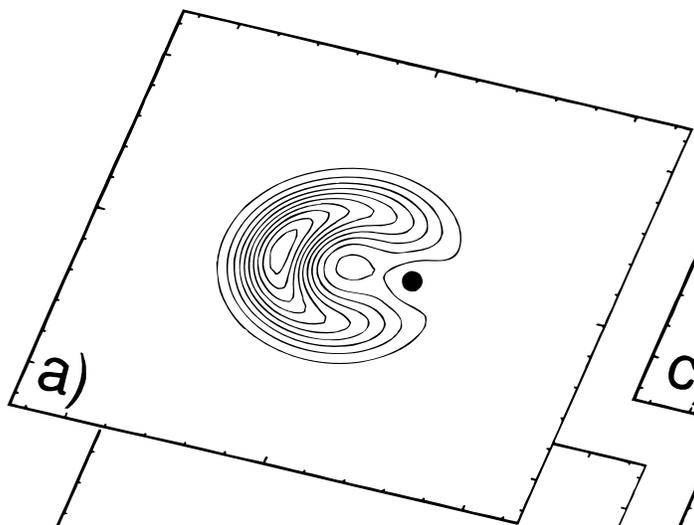
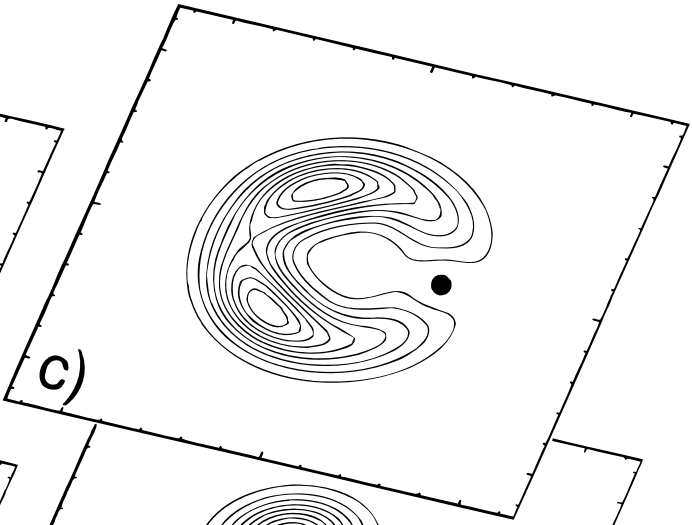
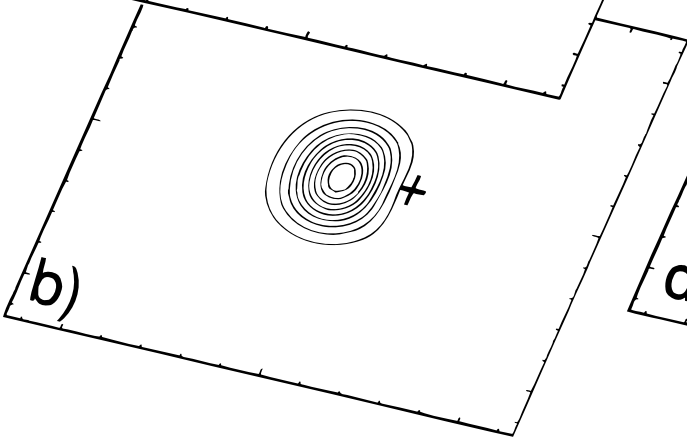
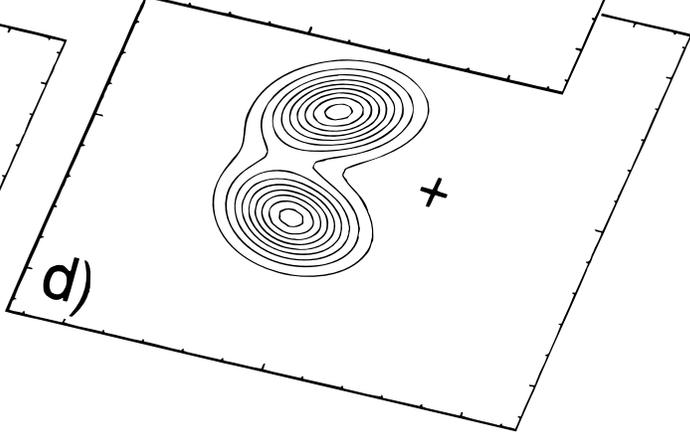
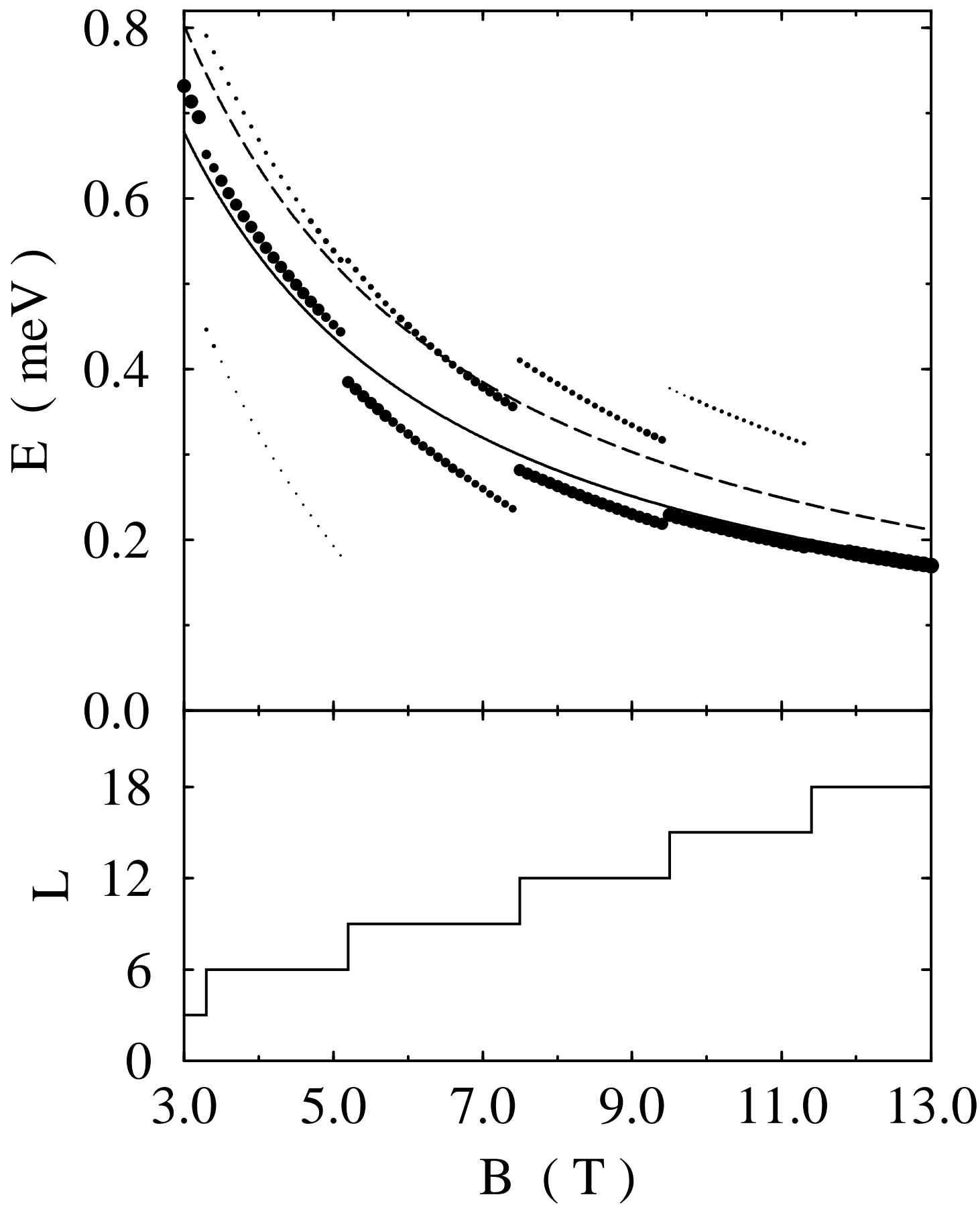

The results of our calculations (Fig.~\ref{fig:fir}) for 
$\Delta_{\rm SAS} = 0.5$ meV shows that 
the spectrum indeed exhibits a series of jumps.  
In a single dot 
the FIR absorption has two branches: the upper branch 
for inter-Landau level transitions and the lower one 
for intra-Landau level transitions. Because we consider only 
the lowest Landau level here we have only calculated the lower
branch  
but we anticipate that the upper branch will exhibit similar jumps.
For comparison, the energy of the lower branch for non-interacting 
electrons given by,
\begin{equation}
\hbar \omega_{\rm single} =
  \frac{\hbar}{2}(\omega_{\rm c}^{2}+4\omega_{0}^{2})^{1/2}-
  \frac{1}{2}\hbar\omega_{\rm c}, 
\end{equation}
is also shown in the figure for 
$\hbar \omega_+ = 2.0$ meV (solid line) and $\hbar \omega_- = 2.2$
meV 
(broken line). It is clear that the coupled dot absorption spectrum 
is not simple like that of a single dot and is split into pieces. 
This means that in the weak magnetic field
region, or in a small total angular momentum region, 
the center-of-mass and relative motions are strongly mixed.  

In particular, 
the jumps in the absorption energy occur at the magnetic fields 
at which the 
total angular momentum changes from one magic number to another.
Thus the ground state transitions should be directly observable in
the FIR absorption spectrum.
The figure also shows that the absorption intensity 
($\propto$ square of the matrix element) is not monotonic. 
For $\Delta_{\rm SAS} = 0.2$ meV 
the FIR absorption spectrum 
is similar, although the 
jumps in energy are smaller than for $\Delta_{\rm SAS} = 0.5$ meV. 

In conclusion, we have found new magic numbers in vertically coupled 
quantum dots and shown that they could be probed experimentally. 

We thank Dr N Bruce for providing results on the electrostatic
potential
of a realistic dot model. This work was supported by the Ministry of
Education, Science and Culture, and the Royal 
Society and the UK Engineering and Physical Sciences Research
Council.

\begin{figure}
  \caption{Ground-state energy against the total angular momentum,
    $L$, 
  in vertically-coupled dots
    with three electrons for 
    the strength of the inter-layer tunneling 
    $\Delta_{\rm SAS}=0.2$ meV (solid line) or 
    $\Delta_{\rm SAS}=0.5$ meV (broken line).  
    Strength of the confinement potential 
    is $\hbar \omega = 2.0$ (2.2) meV for the upper (lower) 
    layer, and the layer separation is $d = 20$ nm.
   Arrows indicate the positions of the cusps.  
    The inset shows 
    a cross section of the charge density in the upper (solid lines) 
    and lower layer (broken lines) 
    against the lateral distance from the 
    center of each dot for $L=5$ (a) and $L=12$ (b).  
    }
  \label{fig:gene}
\end{figure}

\begin{figure}
  \caption{Intralayer (upper panels) and interlayer (lower panels) 
  pair correlation functions, 
    $P({\bf r}, {\bf r}_{0})$, 
    for $L=5$, (a,b) or $L=12$ (c,d).  
   One electron (solid circle) is fixed in the upper layer at
    ${r}_{0}= 16.9$ nm ($L=5$) or 23.5 nm ($L=12$) where the charge
    density has a maximum. 
    The symbol + denotes the projection of the solid circle onto 
    the lower layer.  
    An area with the linear dimension of 128nm is displayed.
    The confinement energies, 
    the layer separation are the same as in Fig.1 with $\Delta_{\rm
      SAS} = 0.2$ meV here.
    }
  \label{fig:prr0l5}
\end{figure}

\end{document}